# An ultra low cost heat source from readily available materials for laboratory and therapeutic applications.

Amit R. Morarka[1*], Aditee C. Joshi[1]

[1]Department of Electronics Science, Savitribai Phule Pune University, Pune- 411007, INDIA.

[*] amitmorarka@gmail.com, amitm@electronics.unipune.ac.in

## Abstract

In most of the under developed nations, access to basic health care is scarce. A heater is an indispensable device for many applications, especially for medical purposes. As such a simple and easy to make heater from readily available materials becomes a valuable equipment to help alleviate the human health problems. An ultra low cost heater from readily available materials like paper/cellulose acetate sheet and a pencil was designed and fabricated. Manually, using high graphite contained pencil; paper/cellulose surface was arbitrarily coated by applying numerous pencil strokes on the surface. Contacts were connected using either a wire or Copper/Aluminum foils over the coated surface. Maximum temperature attended by the heater was around $140^0$C. The main application of this heater was explored in the laboratory and therapeutic uses.

## 1. Introduction

Paper based devices have fascinated significant attention in recent years. In light of this development, another medium of deposition is introduced through pencil sketching. This being a frugal technique although, it has captured prime significance in recent times owing to certain advantages like exclusion of sophisticated facilities of fabrication, no necessity of maintaining specific environmental conditions or using clean rooms. In spite of all this the fabrication process is rudimental yet efficacious. This necessarily indicates ease of implementation on readily available substrates like paper through mechanical abrasions.

Fundamentally, this approach introduces a cost effective, environment-friendly way and provides relatively speedy solution with reference to conventional printing technologies. Pencil, an easy to handle day to day life commodity, is basically a composite made of graphite and kaolin, clay material acting as a binder material [1-3]. Paper has proved to be an effective substrate material owing to its properties like flexible nature and ubiquitous nature, cost effective and compatibility towards deposition techniques. Consequently, by virtue of this superiority many Paper and Pencil on paper devices have been realized on paper substrates in diverse areas of applications such as sensors [4-5], microfluidic devices [6-7], strain gauges [8-9], R-C filters and FET [10] and printed circuit technologies [11].

Pencil drawn chemiresistive sensors have been printed on plastic substrate and studied for strain gauge application [8]. A pencil sketch in form of a planar structure was drawn on a flexible substrate and further tested for various compression and expansion conditions. The sensor was fabricated by using various pencil grades (HB-4B) and sensitivity of all sensor design was compared. A pencil drawn sensor was reported for sensing temperature and pH [4]. A green fabrication method for passive components using a pencil on paper was demonstrated and used in fabricating resistor and capacitor for R-C filter application. In another study, zigzag channels printed through inkjet printing technique were reported for use in paper based micromixers [6]. A paper based microfluidic device was reported using filter paper and studied further for application in acousto-thermal heater and glucose sensing [12]. Several studies report paper based analytical devices in various application. Inkjet printed conductive patterns on paper have been studied for heating characteristics. In addition, this heater was further applied as valve, concentrator and heat source for various chemical reactions in paper based devices. [13]

Pencil trace on paper has proved to be active elements in variety of applications including supercapacitors, piezoresistive sensors, and air electrode in Li-ion battery and electrode materials in UV sensors [5]. Recently, R-C filter and FETs based on pencil drawings on paper were reported along with detailed characterization of morphological, spectroscopic and electrical properties of pencil deposits on paper [10]. Paper based PCBs were also reported by printing electronic circuit elements on different substrates like paper, polyimide, nylon and cellulose [11]. These printed circuits were suitable in applications like low cost RFID, 2-D reconfigurable circuits and consumer electronics. Graphite has been one of the prime important material in semiconducting devices applications. The advent of graphene and more to this 3-D graphene has effectively broadened applications of carbon based devices in diverse domains like catalysis, photocatalytic studies, sensors and energy applications [14]. Pyrolitic graphite is different than pure graphite and its mechanical, thermal and electrical properties are superior to pure graphite. In a recent report, pyrolitic graphite was used in rapid heating and cooling applications. A system based on pyrolitic graphite was designed and characteristics were tested using simulation and experimental approach. The total heating from 50 to 250º C was observed in 2 s and cooling time to 50 C was 8s [15].

Through this manuscript we report the inception of a heater based on pencil strokes on different substrates. The heater is fabricated in a crude manner by mechanical abrasion of graphite composite from soft grade pencils. In turn the fabricated heater was characterized for its I-V characteristics and temperature profiles were studied. The temperature attained with this heater was in range of 40ºC-140ºC with varying input power. The heater was successfully implemented in applications like substrate heating, chemical reactions and has potential to be used in all application demanding higher temperatures in range specified. Additionally, a proof of concept was demonstrated to use graphite composite films in constructing flexible heater on cellulose acetate substrate and further used in therapeutic applications. The entire heater assembly was also tested using battery and the performance was identical as in case of power supply indicative of portability feature. Our results provide an economically viable yet effective heater fabricated through a simple and fast fabrication process without any need of specific instrumentation and ambient conditions.

## 2. Experimental

### 2.1 Fabrication of heater

The substrate materials used in fabrication of heater are mainly readily available notebook paper (average thickness measured using micrometer screw gauge-74.7μm) and cellulose acetate sheets (thickness 100μm) commonly known as OHP (Over Head Transparency) which are used in presentations. Any other make/quality paper will also be sufficed. The pencils used are 9B grade. The paper sheet is cut into different sizes as per heater active area. In our studies, heater area of the size 2" × 1.5" was covered with pencil strokes to lead a conductive surface. For cellulose acetate, a sheet of same dimension was cut and used. The only difference in this approach is that before pencil sketching the surface was made rough using a sand paper for a better adhesion of graphite film to the surface (Figure S1).The pencil strokes are added in a repetitive fashion to add a thin film of graphite composites. After sufficient strokes of 9B pencil there is a visible difference in the paper transparency due to repetitive coating of layers. Also there is metallic luster observed on the surface after pencil strokes are added. Further, the preliminary order of resistance of graphite composite film was noted by placing multimeter probes through copper foils.

These paper films are further used for assembling the heater structure. In our approach we have designed two heater configurations based on the number of heating elements used in parallel with each other. In the first type, the paper/OHP coated with pencil strokes is mounted on plastic substrate (2" × 2") (any other good thermal conductor will also serve the purpose) for mechanical support. The contacts were added by using copper foils having thickness of 100 μm and size 2.7" × 0.8" separated by a specific gap of 1mm. Various gaps (1mm, 3mm and 5mm) were studied for their IV characteristics and corresponding temperatures. Out of all the three gaps, 1mm gap provided minimum resistance for the heater to be operated on low power to get the desired temperature. Subsequently, Alumina substrate was placed above copper foils to provide better thermal conducting medium. Whereas in other type of heater, a paper/OHP substrates coated with graphite composite films was used and copper foils were placed with a specific gap and on top of copper foils another paper substrate of same size coated with graphite composites was placed to get a sandwich structure of heater was prepared. This design was used so as to have a reduced order of resistances under similar applied conditions. To ensure a better contact with the paper, all this assembly was held with various binder clips for a perfect contact between copper foils and graphite composites film. The entire assembly was tested under various applied input power conditions. Figure 1 shows the cross sectional view of sequence of layers in the structure of heater.

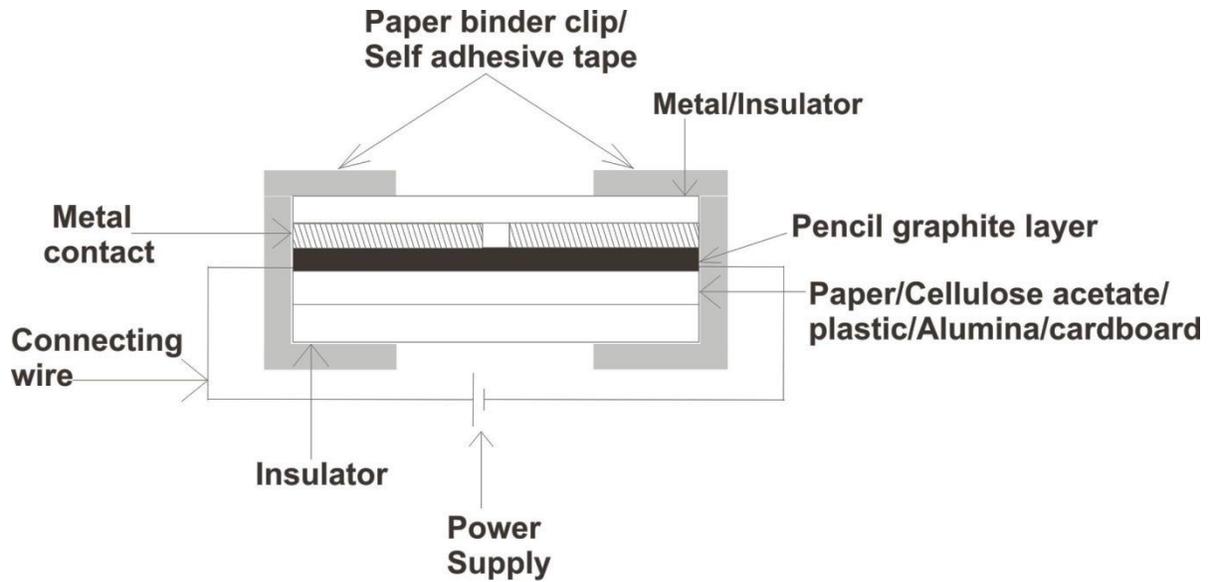

Figure 1 Basic structure of heater assembly.

## 2.2 Design of heater with wires as contact electrodes

The effect of copper foil as electrode material was also studied. For comparison wires were attached instead of copper foils as contact electrode on the graphite coated region. Figure 2 shows structure of heater with copper foils (figure 2a) and one with wires (figure 2b) used as contacts. The effect of contact electrodes on heat transfer rate was analyzed.

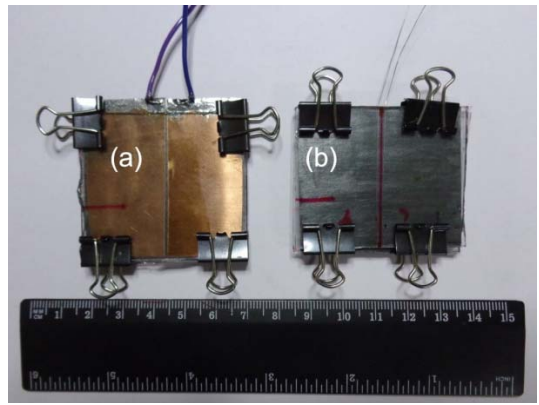

Figure 2 Image from the top of the heater assembly a) Copper foils as contact electrodes.
b) Wires as contact electrodes. In the (b) image, a red line marked on the top glass surface runs parallel to the wires which are 1mm apart. The horizontal line marked on both the heater surfaces signifies the place where the PT-100 probe was located for recording the temperatures.

## 2.3 Modification of heater structure for temperature gradient study

The temperature profile of the heater was tested by applying a voltage sweep in a step of 50mV and checking corresponding current through the heater. The temperature of heater surface was monitored by measuring resistance of a calibrated PT-100 sensor placed on the alumina substrate. Care was taken to position PT-100 in a close proximity to heater surface near the active heating area to avoid any errors during temperature monitoring. I-V characteristics of heater design were repeatedly recorded for a given heater structure and other same designs to ensure repeatability and reproducibility in a given design.

In order to study temperature gradient across heater surface temperature sensors were placed on three positions and gradient was also monitored through modification in heater design. The heater assembly was designed with multiple numbers of electrodes contributing to more number of resistances across same heater surface. The assembly can be used for a single graphite composite film or a sandwich structure having two graphite composite films. In case of the two graphite films, the electrodes are sandwiched in between the films. The temperature profile for this design was tested by placing three temperature sensors across two edges and one in middle section of the heater. The structure of the heater is as shown in Figure 3a; it consists of three copper foils. A gap of 1mm was maintained between two consecutive foils. The heater dimensions designed were 3.2" × 1.6". Copper foils used as the electrodes near the edges were having width 0.4". The width of middle copper foil was 2.3".

In the same approach another design was tested in which a heater same as shown in figure 1 was fabricated but with a blank area left in middle region of paper as seen in Figure 3b. The rest of the paper was coated with graphite composite film (Figure S2). In particular two shapes viz. square and rectangle was designed and gradient was studied as shown in figure 3(a, b). The concept can be adopted for different shapes and number of blank regions on the paper. The temperature gradient was monitored at three different points across edges and middle part of blank region.

## 2.4 Design of a flexible heater

Subsequently, a flexible heater was designed and tested. The heater element was mounted on a 100μm aluminum foil using single sided self-adhesive tape which provided a flexible support. The aluminum foil helped in providing a uniform distribution of temperature over the area of application. In turn this was applied for therapeutic application.

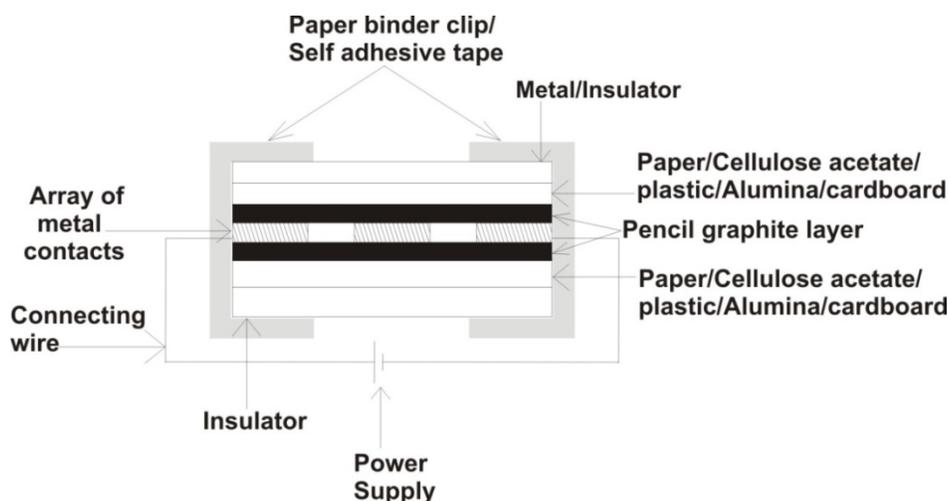

Figure 3a Structure of heater assembly with multiple electrodes.

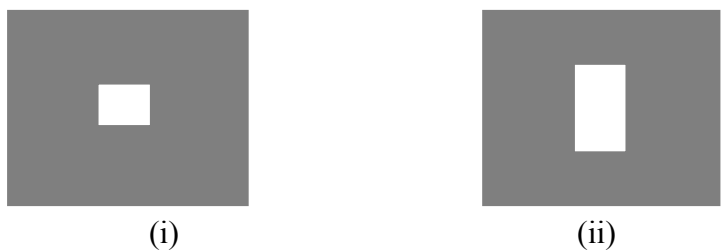

(i)                       (ii)

Figure 3b Structure of heater assembly with blank regions i) Square shaped blank region. ii) Rectangular shaped blank region

## 3 Result and Discussion

### 3.1 I-V characteristics of graphite composite films

The electrical characteristic of the heater was studied by I-V characteristics as indicated in figure 4. The typical value of resistance for the heater was in the range from 5-20 Ω. The I-V curve exhibited almost linear nature in the initial region while in the other regime as the voltage increased the current increase was very rapid and more or less the I-V curve was non-linear due to swift in overall conductivity of graphite composite film with reference to film surface temperature as reported earlier [10]. In turn this characteristic of increased conductivity with temperature for graphite composite or pencil trace is favorable in our heater application, as it helps in increasing the heating rate at a given bias. Subsequently, we are getting higher temperature at a relatively lower power inputs.

We have studied the heating characteristic for graphite composite trace on paper (figure 4a) as well as cellulose acetate substrate (figure 4b). I-V characteristic for both the heaters exhibited a similar linear trend for applied voltage. The temperature of the heater surface was logged through a calibrated PT-100. The temperature for a specific value of PT-100 resistance was obtained through a calibration table available for PT-100. The resistance of PT-100 and its corresponding temperature values were recorded. It was observed that when voltage is increased there is increasing current through the heater that heats up the surface in more or less linear fashion. The final temperature achieved through the heating was just over 100º C. The time taken by the heater to achieve this temperature for given current and voltage values was of order of few minutes.

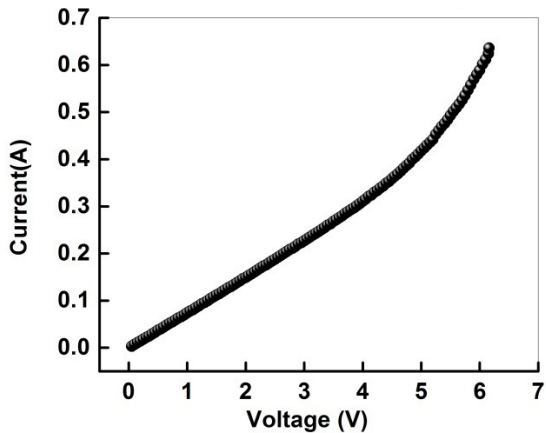
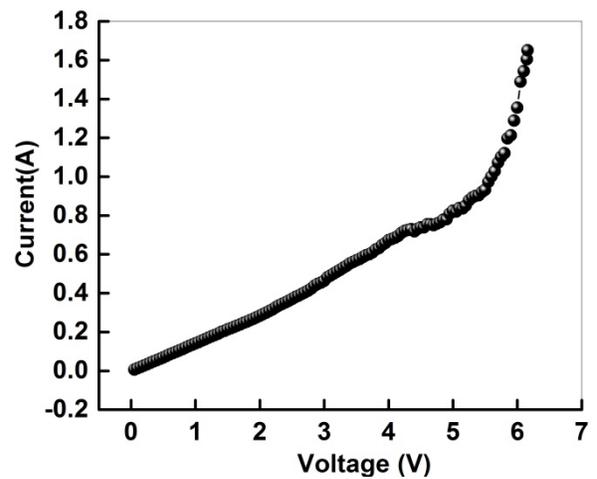

Figure 4a) I-V characteristic of heater on paper substrate.

Figure 4b) I-V characteristic of heater on cellulose substrate.

## 3.2 Comparison of copper foil and wires as contact electrodes

We have also studied the comparison of copper foils and wires as contact electrodes. Figure 5 shows rate of change of temperature at a specific location (same in both designs). The rate of change of temperature was calculated in both the designs by calculating slope of linear regions in both curves. It was observed that rate of change of temperature was quite rapid in case of copper foils (data shown in red diamonds) (6 mΩ/s) and for wires as electrodes (data shown in black circles) it was quite slow (3mΩ/s). This might be due to higher surface area of copper foils available for thermal conduction across surface. Nevertheless, wires as electrode materials could also be useful in applications where rate of heating is of no concern.

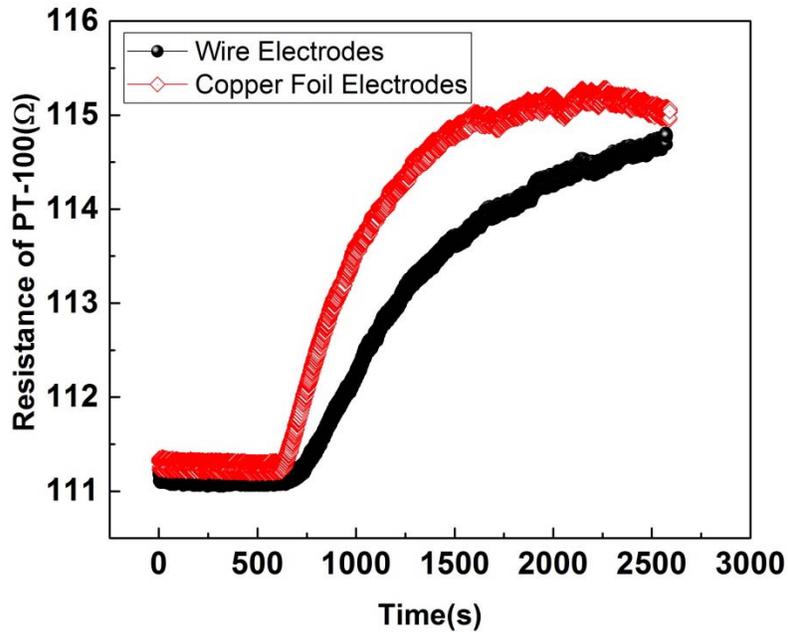

Figure 5 Comparison of heating rate for copper foils (red diamonds) and wires (black circles) as contact electrodes

### 3.3 Temperature-Time profile of heater

The total time duration required for approaching 100º C in above I-V characteristic was also monitored. Figure 6 shows the increase in temperature with time for I-V values analogous to range as specified above. The maximum temperature obtained was more than 100º C, however the time required for getting 100º C was 3 minutes from room temperature value. Over the set of observations for different heaters the 100º C temperature was achieved in maximum 5 minutes.

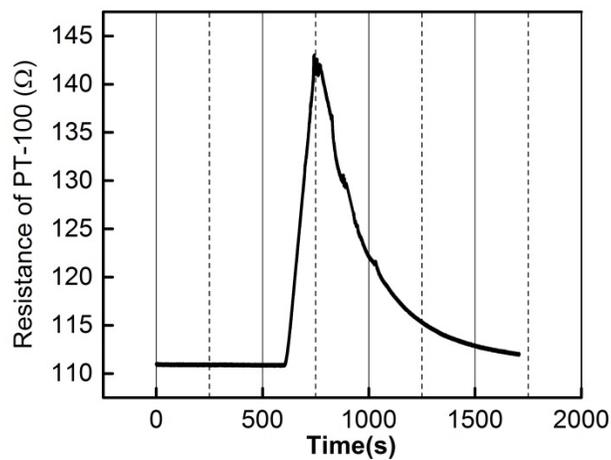

Figure 6 Temperature and time profile for heater. The heater was kept at ambient temperature for some time. Heating was started after the PT-100 attained a stable value.

### 3.4 Repeatability and reusability characteristics

The reusability of the heater greatly depends upon the final temperature for which it is to be used. Around low temperatures in the range of 40-60ºC the heater can be used multiple times. But at the higher operating temperatures, it is useful for no more than 2-3 times reliably. We can speculate some explanations based on two major conditions. The first one is related to the wrinkles on the surface of the paper. While coating the surface of the paper with the pencil strokes, accidently the paper undergoes bending. If the bending causes crease in the paper (or the paper already has it), the local current and hence the temperature distribution alters significantly as compared to the rest of the coated surface. If the altered distribution is towards increased current, corresponding temperature also increases. This causes pockets of high temperature region. These localized high temperature zones are amongst the root cause towards the local burning of the paper. This degradation of the paper is what limits the reusability of the heater. Similarly in the 2$^{nd}$ case, since the pencil strokes are not uniformly distributed while coating the paper, local resistance variations are prominent on the graphite composite coated paper. If the electrodes cover a certain region on the paper which has at that point very low resistance, the current flow through this region will be significantly high. This causes enough high temperatures for the paper to undergo local burning. Even with such limitations on the reusability of the heater, its low cost and simple fabrication process makes it an easy to build do it yourself heating source for various applications.

### 3.5 Applications of heater

Further, to study application of this heater we performed various processes and chemical reactions that depend on elevated temperature ranges. Figure 7 depicts changes in different reactions at room temperature and elevated temperature. We have placed silica gel saturated with humidity on top surface of heater (Aluminum plate) (figure 7a) and after heating it showed a drastic change in its color from transparent to opaque white (figure 7b). It takes around 60 º C temperature to change the color of silica gel, and spontaneous change in color indicates that temperature achieved on heater surface is of similar order. We have also monitored the dehydration of cupric sulfate, which happens at higher temperature. In our experiment we added a small amount of cupric sulfate powder on top plate of heater at room temperature (figure 7c) and after heating it showed immediate change in color from its original color and turned into white color powder (figure 7d). Additionally, to crosscheck the temperature value approaching to 100ºC, we have put a drop of water on top plate of heater and it was immediately evaporated in few minutes. Consequently this heater can be effectively useful in substrate heater kind of applications as well, we have tested this for a glass substrate with an ink droplet (figure 7e) and after few minutes time interval a dried film of ink was observed on the surface of glass substrate (figure 7f). $FeCl_3$ film was also obtained on glass substrate using a solution of $FeCl_3$ in a similar manner (figure 7 g and h).

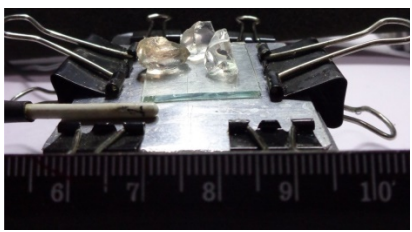

Figure 7 (a) Saturated Silica gel added to preheated heater surface.

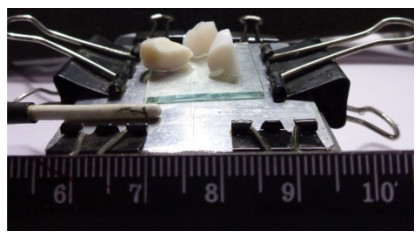

Figure 7 (b) Silica gel after water evaporation.

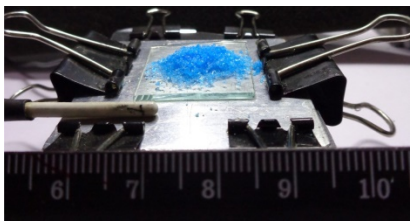

Figure 7 (c) Cupric Sulphate crystals added to preheated heater surface.

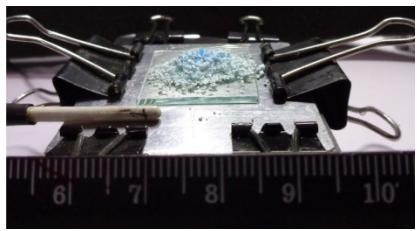

Figure 7 (d) Cupric Sulphate crystals after heating.

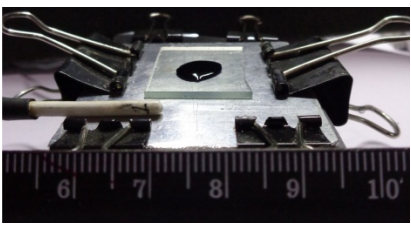

Figure 7 (e) Ink drop at solution added to preheated heater surface.

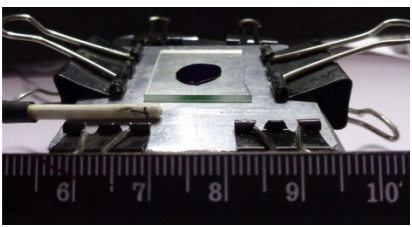

Figure 7 (f) Ink drop converted in a film after water evaporation.

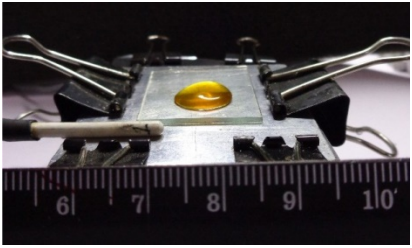
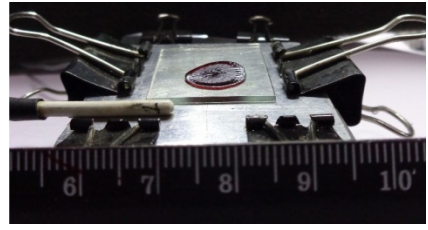

Figure 7 (g) FeCl$_3$ solution added to preheated heater surface.

Figure 7 (h) FeCl$_3$ film after water evaporation.

### 3.6 Temperature gradient study

#### 3.6.1 Heater design with multiple electrodes.

The temperature profile for heater with multiple electrodes assembly was tested. The temperature was monitored at three different regions. Figure 8 shows graph of temperature variations across three different positions of sensors. It was observed that the temperature was higher near region where sensor-3 was placed. The effective difference in the temperatures across three sensors was calculated. It was noted that a gradient of ~7ºC was developed between third sensor and either of first or second sensor. This can be significant in certain studies where various temperature zones are necessary and can be adopted in such applications. The temperature gradients and corresponding temperature zones can be increased based on dimensions and number of electrodes. Moreover, the resistance of heater can also be tailored by varying the gaps between individual electrodes and subsequent gradient can be obtained.

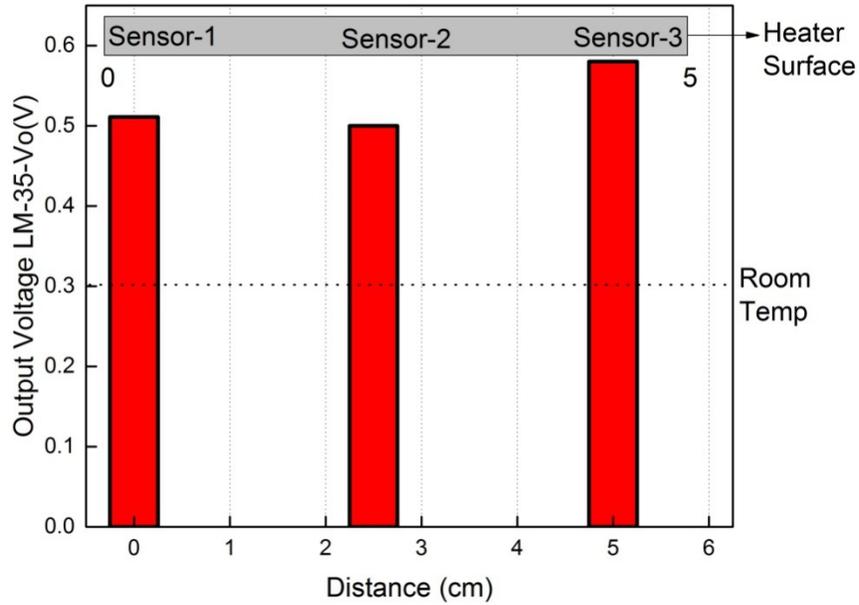

Figure 8 Temperature gradient for heater assembly with multiple electrodes.

### 3.6.2 Heater design with blank regions.

In the heater design with blank regions on the paper, gradient substantially changed across the blank space. The distribution of temperature across heater with blank regions is shown in figure 9. It was observed that as area of blank region increases the gradient monitored across edges and in middle section of blank region also increases in more or less linear fashion. The maximum gradient observed over the rectangular blank region was in between sensor 1 and sensor 3 was around 30ºC at elevated temperatures. On an average 10-20ºC gradient was recorded in between sensors 1-2 and 2-3 respectively. Based on the observations of temperature gradients on the heater with copper foils as the electrodes, we can extrapolate that, with the same heater but with wires as the electrodes instead of copper foil, temperature gradients higher than the above reported values can be achieved.

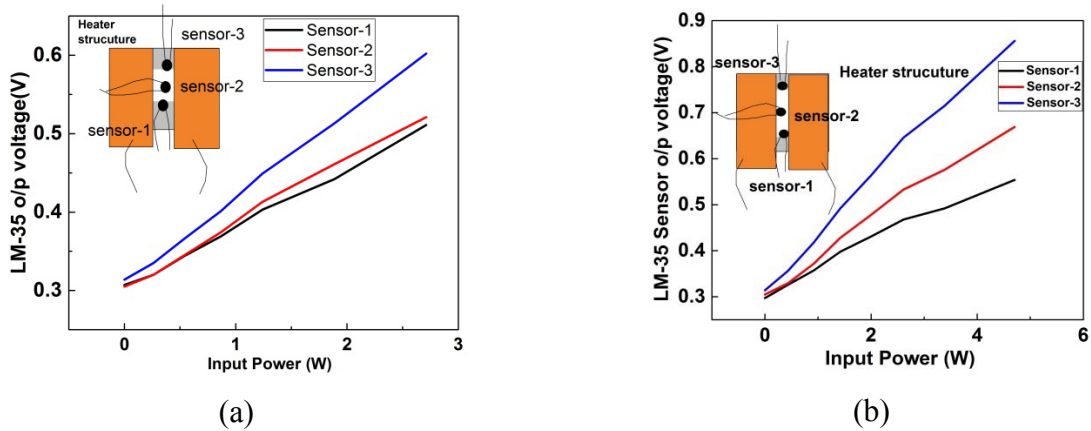

(a)                                (b)

Figure 9 Temperature gradient for heater with blank regions: (a) Square shaped blank region. (b) Rectangular shaped blank region.

### 3.7 Design of a flexible heater

After the successful evidence of heater in application for chemical reactions as well as substrate heating or baking applications, this heater was also tested for its application in therapeutic domain as in for the biomedical purpose. As outlined in the experimental section, the heater was flexible in nature so as to have ease of application over the human body parts. This was subjected to linearly increasing voltages and resultant current was noted. It was observed that the straight design of the heater did not show any significant base conductivity due to improper contact between pencil trace and contact electrodes. Further, this heater was mounted on a cylindrical cardboard support (ID= 4.6 cm and wall thickness= 0.1 cm) using commercially available wrist band. The tightening of the wrist band provided enough pressure over the heater to obtain heater resistance around 26 Ω. Then the heater was mounted on 90 μm Aluminum foil and fixed using self-adhesive tapes. Electrical contacts through the wires were established (Figure S3). The heater arrangement was then mounted on a human forearm using a wrist band. The band was tightened enough to provide good ohmic contacts with electrical connector and heater due to its pressure. The pressure was adjusted in a way that it would not cause any blood flow inhibition in the arm. To monitor the temperature PT-100 was inserted between skin and the heater. The heater voltage was taken upto 5V slowly and the current and temperature rise was monitored. The final temperature was set to 37ºC as shown by PT-100 resistance (114.3Ω). At this temperature, the heating effect was significantly large and uniformly distributed over the wrapped surface. It was within the limit of the human comfort. The heater assembly mounted on human forearm is shown in figure 10.

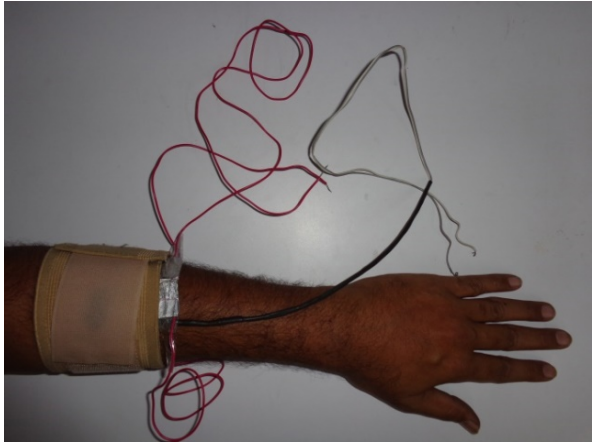 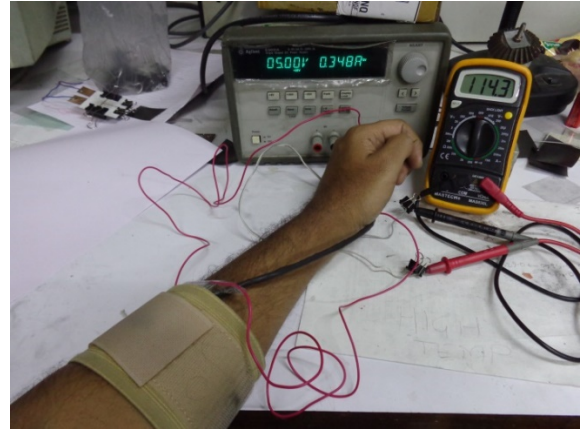

Figure 10 (a) Heater assembly wound on forearm with contacts for PT-100 (white and gray wires inside a thick black sleeve) and contacts for power supply (red wires).

Figure 10 (b) Heater temperature achieved at given power. The multimeter displays the resistance of PT-100 at 37ºC.

**3.8 Portability study**

In order to study the portability of the device, we have checked temperature response of one heater using a 5V battery (power bank of 5V, 2A). From figure 11(a) it can be seen that when the heater is disconnected from battery it displays ambient temperature whereas after connecting the battery the temperature increased (figure 11b). This shows prospect of using a battery for operating a heater and makes it portable for remote applications.

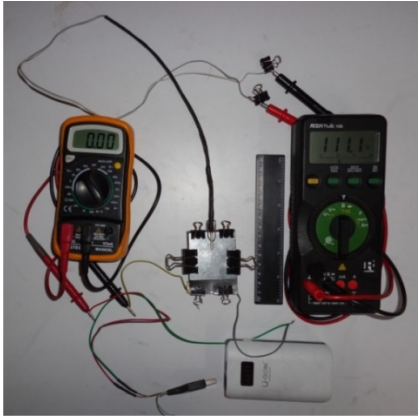 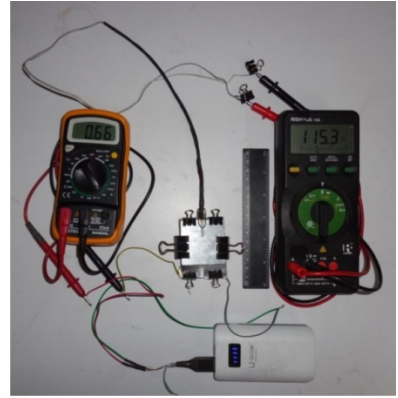

Figure 11(a) 5V battery switched OFF. Temperature of the heater 28$^0$C

Figure 11(b) 5V battery switched ON. Temperature of 40$^0$C was attended by the heater in approximately 3 minutes.

## 4.0 Conclusion:

We have developed a new fabrication methodology for paper based heaters using pencil on paper approach. The fabrication process is simple, fast and involves pencil traces on paper that yields a device. In turn this demonstrates simple, flexible, cost effective and portable heater useful in variety of applications. The heater device was successfully demonstrated as a heating application in chemical reaction, substrate drying application and baking purpose in fabrication processes. Additionally, this was significant in therapeutic application for biomedical domain.

**Acknowledgement**

The authors wish to thank Prof. S. AnanthaKrishnan, Adjunct Professor and Raja Ramanna Fellow, Department of Electronics Science, Savitribai Phule Pune University (SPPU) for providing his laboratory and the resources for this work. Authors are thankful towards Mr. Somnath Bhopale, Physics department, (SPPU) for his motivational discussions for this work.

# Supplementary Information

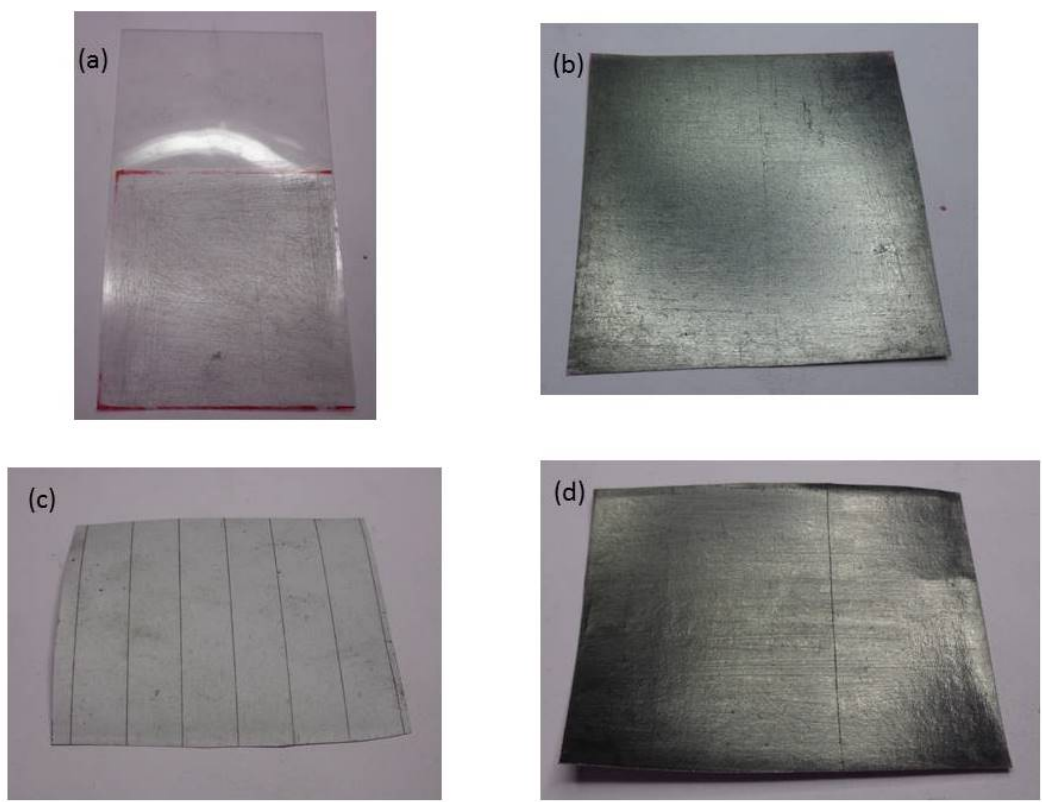

Figure S1: Images of the paper and OHP sheet. (a) OHP sheet with and without roughing the surface (b) After coating with pencil strokes (c) Plane paper sheet (d) After coating with pencil strokes.

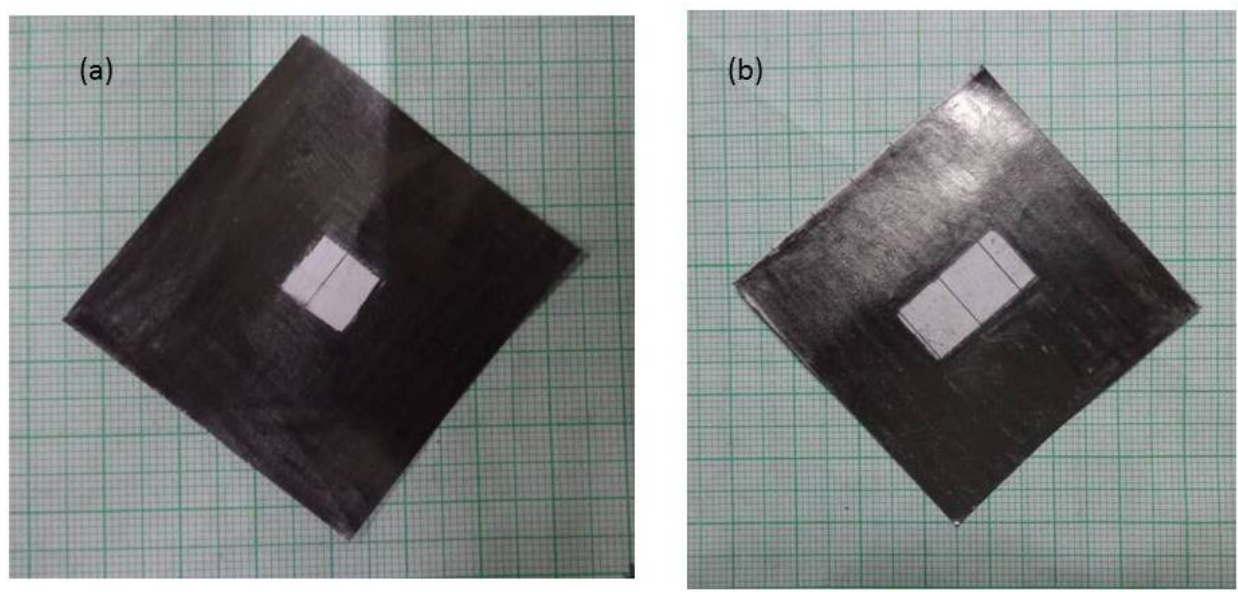

Figure S2: Images of the paper coated with pencil strokes having the blank area of (a) Square shape and (b) Rectangular shape.

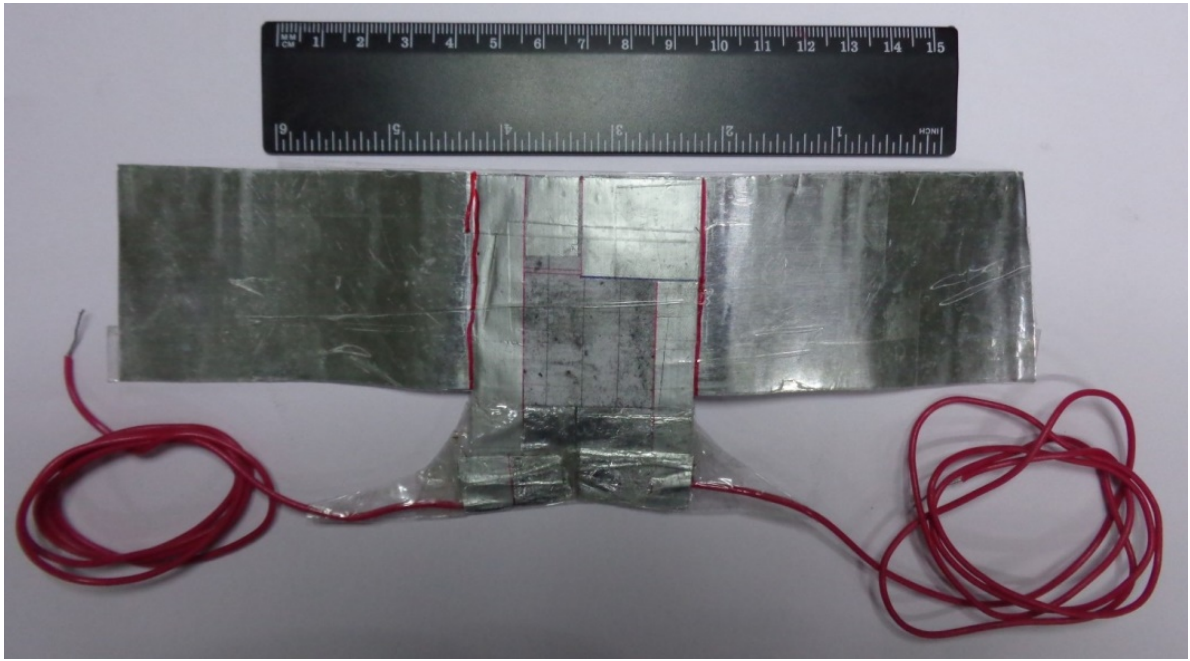

Figure S3: Image of the final heater fabricated using pencil strokes on paper for therapeutic application. Aluminum and wires were used for making the electrical contacts. Heater was mounted on aluminum foil having thickness of 100μm for uniform heat dissipation over the larger area